# Introducing the PIT-plot – a new tool in the portfolio manager's toolkit

Stig Johan Wiklund, Magnus Ytterstad

Captario
Göteborg, Sweden

E-mail: stig-johan.wiklund@captario.com

## Abstract

Project portfolio management is an essential process for organizations aiming to optimize the value of their R&D investments. In this article, we introduce a new tool designed to support the prioritization of projects within project portfolio management. We label this tool the PIT-plot, an acronym for Project Impact Tornado plot, with reference to the similarity to the Tornado plot often used for sensitivity analyses. Many traditional practices in portfolio management focus on the properties of the projects available to the portfolio. We are with the PIT-plot changing the perspective and focus not on the properties of the projects themselves, but on the impact that the projects may have on the portfolio. This enables the strategic portfolio management to identify and focus on the projects of largest impact to the portfolio, either for the purpose of risk mitigation or for the purpose of value-adding efforts.

## 1. Introduction

Project portfolio management (PPM) is an essential process for organizations aiming to optimize the value of their R&D investments. Numerous methods, both qualitative and quantitative, have been proposed and used for the management and selection of projects in the portfolio (Raada et al. 2020, Lerch and Spieth 2013, Mohagheghi et al. 2020, Vieira et al.2024), with some implementations employing formal optimization procedures (Sampath et al. 2022, Farid et al. 2021). Other authors focus on the organizational and human behaviors in the PPM process (Martinsuo et al. 2024, Schiffels et al. 2018). One of the key challenges in PPM is understanding and managing the uncertainties inherent in each project within the portfolio (Hu and Szmerekovsky 2017). Quantitative tools such as sensitivity analysis becomes invaluable in these evaluations (Eschenbach 1992).

As a contribution to enhancing the decision-making toolkit of portfolio managers, we will in this paper introduce a novel analysis tool: the Project Impact Tornado plot (PIT-plot). The PIT-plot is a graphical representation of the quantitative impact of each project on an aggregated portfolio metric of interest. While it shares some visual similarities with the traditional Tornado sensitivity analysis plot (Eschenbach 1992), the PIT-plot serves a different purpose. It is specifically tailored to illustrate the investment efficiency of projects within the portfolio, providing a more focused perspective on project risks and returns in the context of the broader portfolio.

To set the stage for understanding the PIT-plot, we will first provide a brief overview of the core principles of project portfolio management and the role of sensitivity analysis in this context. We will then describe the details of the PIT-plot, illustrating its application and benefits within the pharmaceutical industry. Although our discussion is grounded in the specific demands of drug development, the concepts and insights we present are broadly applicable to other industries where portfolio management and sensitivity analysis play pivotal roles in strategic decision-making.

## 1.1 Project Portfolio management

Project portfolio management (PPM) is used in organizations to manage and align a collection of projects with their broader business objectives. Unlike traditional project management, which focuses on executing individual projects efficiently, PPM emphasizes the optimization of a portfolio of projects to maximize overall value and achieve strategic goals.

In the pharmaceutical industry, all mid- and large-sized companies maintain a portfolio of several drug development projects, and PPM has been used as an important strategic process for several decades (Tiggemann et al., 1998). These projects may target diseases in different therapeutic areas, and they may vary in scope, scale, and strategic importance. The objective of PPM is to enable the evaluation, prioritization, and selection of projects based on their potential return on investment, alignment with organizational strategy, and resource availability.

In the process of portfolio management, information and data from a variety of sources need to be synthesized. Examples include cost estimates, sales forecasts, risk assessments and resource requirements. Numerous tools and methods, both quantitative and qualitative, are available and used by portfolio managers. The purpose of this article is to provide a new tool to the toolkit of professionals active in PPM.

## 1.2 Sensitivity analysis and the Tornado plot

Tornado plots (or Tornado diagrams) are often used for sensitivity analysis, enabling the comparison and visualization of the relative importance of variables in a quantitative model. To create the Tornado plot, the outcome of a model is evaluated in an iterative procedure for each of the variables in the model. The variable currently evaluated in each step is modelled as having an uncertain value while all other variables are held fixed at baseline values. The uncertainty is typically represented by low, base and high values of the variable.

The plot is created as a bar chart, in which the bars are placed horizontally, and the variables of interest are placed vertically. The lengths of the bars are given, for each variable, by the difference between the model outcome for the low, base and high scenarios. The variables are sorted and placed so that the most influential variables appear at the top of the plot, and the least influential at the bottom. This implies that the longest bars are at the top and the shortest bars at the bottom, giving the plot the shape of a funnel or tornado, which has given the plot its name. A simple illustrative example of a Tornado plot is given in an Appendix.

The Tornado plot is frequently used as a tool for sensitivity analysis, as it illustrates the sensitivity of a metric of interest to changes in a set of input variables (Eschenbach 1992). In the setting of a drug development project, the Tornado plot may be used to illustrate how e.g. the expected net present value, eNPV, of a drug project would be affected by changes in parameters like development costs, launch timing, price, market share, etc. The Tornado plot may serve as a useful tool to illustrate which opportunities and risks in a project might have the largest impact and hence being candidates for increased attention.

## 2. Introducing the PIT-plot

The traditional Tornado plot was introduced in the previous section, and its usefulness in project sensitivity analysis was outlined. In the portfolio management setting, however, the over-arching questions are different. Key questions then deal with the composition of projects to build an optimal portfolio of projects. Large and mid-sized pharmaceutical companies do often have access to more development opportunities than they could fund with the available budget. Using classical sensitivity analysis, and the Tornado plot, is not particularly relevant to this situation. Any single parameter in a given project is not likely to have a major impact on the portfolio. The thousands of potential parameters involved in quantitatively representing a portfolio may create a multitude of estimated sensitivities that are difficult to interpret and each may represent only marginal impacts. In situations where the analysis accounts for uncertainty, and estimates are obtained through Monte Carlo simulations, the results may further be blurred by random noise.

We do instead suggest that the sensitivity analysis at the portfolio level is primarily based on the impact of changes in the portfolio composition, not on each of the parameters that might contribute to the value of each of the projects. The inclusion or exclusion of a project will have an impact on the portfolio metric of interest, which could be estimated by an appropriate quantitative model. Quantifying this impact of a project on portfolio metrics will provide the basis for a relevant sensitivity analysis in support of portfolio management. Building on the similarities with the traditional sensitivity analysis for variables in a project model, and the corresponding Tornado plot, we introduce the Project Impact Tornado, i.e. the PIT-plot, as a tool for sensitivity analysis of the impact of projects in a portfolio management setting.

### 2.1 Notation and definition – general case

We will in this section give the generic definition of the components of a PIT-plot, applicable to any portfolio metric for which the project sensitivity is of interest. A schematic illustration of the PIT-plot and its components is given in Figure 1.

The key components of the plot are the following:
$M_P$      Risk-adjusted metric of interest calculated for the entire portfolio
$M_i$      Risk-adjusted metric of interest calculated for project $i$
$M_P^{(i)}$      Risk-adjusted metric of interest calculated for the portfolio, when project $i$ is excluded
$\widetilde{M}_P^{(i)}$      Metric of interest calculated for the entire portfolio, conditional on project $i$ being successful and generating revenue $\Delta^{(i)} = M_P^{(i)} - M_P$    Length of the PIT-plot bar for project *i*, representing the impact on the metric of interest from excluding project *i*. We will be referring to this as the "Exclusion bar".

$\widetilde{\Delta}^{(i)} = \widetilde{M}_P^{(i)} - M_P$    Length of the PIT-plot bar for project *i*, representing the impact on the metric of interest if a successful outcome of project *i* could be guaranteed. We will be referring to this as the "Success bar".

The PIT-plot is generated by producing a horizontal bar chart, with two bars for each project in the portfolio. The bars extend from a vertical center line of the plot, and the length of the bars are given by $\Delta^{(i)}$ and $\widetilde{\Delta}^{(i)}$, respectively. The projects are ordered by their value of $\Delta^{(i)}$, with the order being from the smallest (i.e. typically the most negative) to the largest.

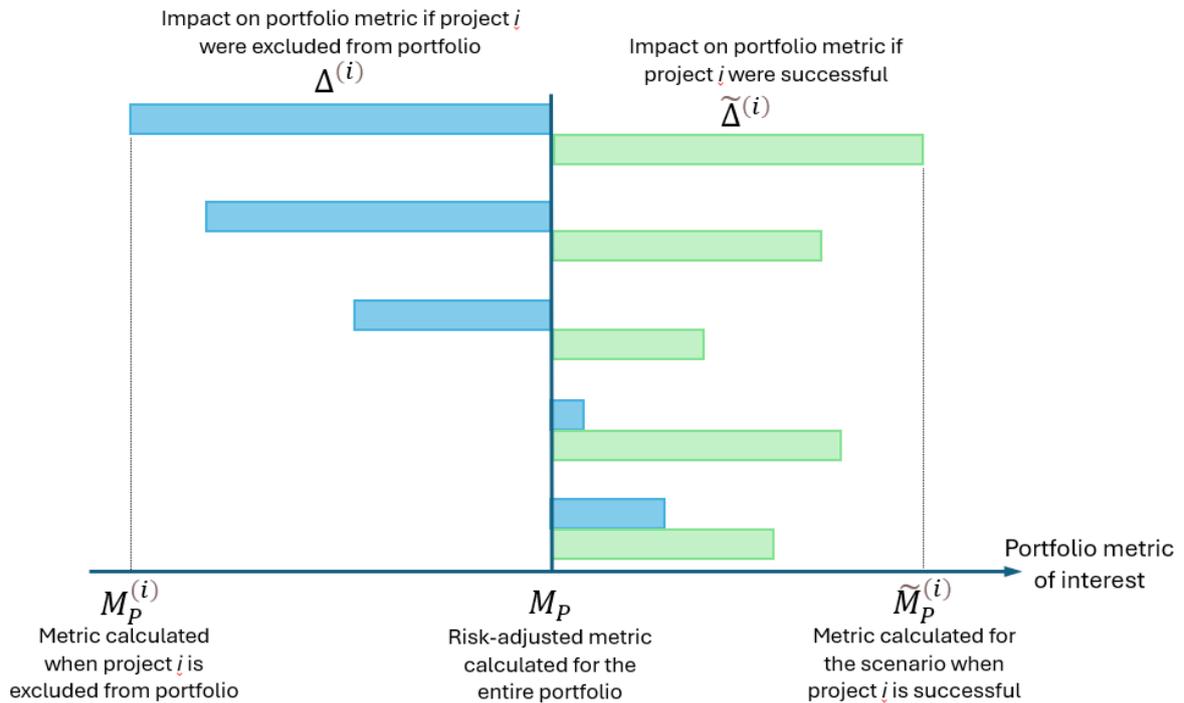

*Figure 1. A schematic illustration of the components of a PIT-plot*

## 2.2 Notation and definition – metrics based on simulation of future cash flow

We will in this section give a more detailed account of how the components of a PIT-plot would be obtained in the situation where the metric of interest is based on forecasts of future cash flows. We will also give an extension to the situation where the metric, and the underlying cash flows, are obtained through a Monte Carlo simulation.

The procedure will be illustrated for a metric referred to as the Productivity Index, PI, which is a return-on-investment type metric defined as

$$PI = \frac{R - C}{C}$$

where $R$ is the expected revenue, and $C$ is the expected development cost required to get the project to the market.

Assuming that the revenue and costs are forecasted as cash flows at future time points, $t$, we have for each project
$R_i = \sum_t r_{it}$ and $C_i = \sum_t c_{it}$
When the cash flow data are obtained from a Monte Carlo simulation, the corresponding values are obtained as
$R_i = J^{-1} \sum_j \sum_t r_{ijt}$ and $C_i = J^{-1} \sum_j \sum_t c_{ijt}$
where $J$ is the number of iterations in the simulation.

Calculating the metric for the entire portfolio implies summing over the projects as
$R_P = \sum_i R_i$ and $C_P = \sum_i C_i$
and calculating the $PI_P$ according to the definition of the Productivity Index above. This would correspond to $M_P$ in the generic notation outlined in the previous section and represent the value of the mid-line of the PIT-plot.

Excluding project $i$ and summing over the remaining projects provides the revenue and development costs as
$R_P^{(i)} = \sum_{i^* \neq i} R_{i^*}$ and $C_P^{(i)} = \sum_{i^* \neq i} C_{i^*}$
Calculating the $PI$ for each project using the values $R_P^{(i)}$ and $C_P^{(i)}$, yields $PI_P^{(i)}$ corresponding to $M_P^{(i)}$ in the generic notation of the PIT-plot.

We will assume here that the Monte Carlo simulation includes iterations for scenarios where each project will fail, as well as scenarios where each project will be successfully launched to the market and generate revenue. Let $\tilde{j}_i$ represent the iterations where project $i$ is successful, we can take the risk adjusted revenue and development costs for the other projects and add the conditional revenue and development costs for the scenario that project $i$ is successful.
$\tilde{R}^{(i)} = \sum_{i^* \neq i} R_{i^*} + \tilde{J}^{-1} \sum_{j \in \tilde{j}_i} \sum_t r_{ijt}$ and $\tilde{C}^{(i)} = \sum_{i^* \neq i} C_{i^*} + \tilde{J}^{-1} \sum_{j \in \tilde{j}_i} \sum_t c_{ijt}$
Calculating the PI for each project using the values $\tilde{R}_P^{(i)}$ and $\tilde{D}_P^{(i)}$, yields $\widetilde{PI}_P^{(i)}$ corresponding to $\widetilde{M}_P^{(i)}$ in the generic notation of the PIT-plot.

The length of the exclusion bars and success bars, respectively, of the PIT-plot would in the example with the Productivity Index, for each project, be given by
- Exclusion bar: $\Delta^{(i)} = PI_P^{(i)} - PI_P$
- Success bar: $\tilde{\Delta}^{(i)} = \widetilde{PI}_P^{(i)} - PI_P$

It may be noted that the calculation of financial metrics for project and portfolio evaluation is often based on values that are discounted to a present value. Nominal forecasts for a cash flow are then discounted using a present value conversion factor $V_t = 1/(1 + q)^t$, where $q$ is the discount rate. The discounted cash flows are
$r_{ijt} = V_t \bar{r}_{ijt}$ and $c_{ijt} = V_t \bar{c}_{ijt}$
where $\bar{r}_{ijt}$ and $\bar{c}_{ijt}$ are the nominal cash flow forecasts, and the discounted cash flows could be used as illustrated in the calculations above.

While this section has illustrated the explicit calculation of a specific financial metric, the Productivity Index, it should be reiterated that the PIT-plot concept could be applied to any quantitative metric of interest.

*2.3 Interpretation and usage*

We will in this section provide some examples of how the findings of a PIT-plot could be interpreted and used in project portfolio assessment and prioritization.

*Identifying candidates for project termination*

Projects at the lower end of the PIT-plot may be identified as prime candidates for project termination in the case where there is not enough budget to fund all available projects, or where alternative investment options are available. For some metrics of interest, in particular relative measures like Return on Investment, Internal Rate of Return, etc., excluding projects at the lower part of the PIT-plot will have a positive impact on the portfolio-wide measure. This corresponds to the "exclusion bar" of the PIT-plot extending to the positive side of the plot's vertical center line.

*Identifying project candidates for additional focus and risk mitigation*

Projects having a large positive 'success bar' in the PIT-plot may be identified as candidates for efforts in risk mitigation and additional measures for increasing the project's success probability. The large positive size of the success bar implies that there is much to gain in terms of positive impact on the portfolio metric of interest, should the project eventually be successful. Hence there is a large upside that warrants investments and efforts to increase the probability of success for such projects.

*Identifying projects for which continuing according to current plans might be prudent*

The previous paragraphs indicated how the PIT-plot might help identify projects for which alternative courses of action could be considered. However, for some projects the "exclusion bar" may show a large negative impact and the "success bar" show a small or moderate positive impact. These projects are not candidates for termination, as the exclusion of these projects would have a large negative impact on the portfolio metric. However, any risk mitigation to increase success probabilities will not have a major positive impact on the portfolio, hence the conclusion might be to allow these projects to proceed according to current plans.

# 3. Illustrating example

We will in this section illustrate the use of a PIT-plot on a small portfolio of ten projects. The illustrating example has taken inspiration from a mid-sized pharmaceutical company, but composition and numbers have been substantially modified. For the illustrating example, the calculation of cash flows are based on applying a generic model for a drug development project (cf Wiklund, 2019), as schematically illustrated in Figure 2. Key attributes of this model are the duration of each phase, $D_h$, and the cost incurred at each phase, $C_h$, for $h \in \{1,2,3,r\}$. The probability that the project successfully proceeds from Phase $h$ to the next phase is denoted $P_h$. The revenue from the project is incurred in the market phase and is denoted $R_m$. Empirical data used for the illustrating example are summarized in Table 1.

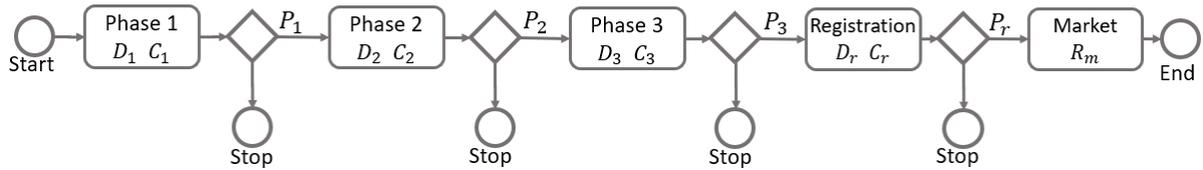

*Figure 2. Schematic illustration of a generic drug development process.*

*Table 1. Input parameter values for the projects included in the illustrating example.*

|  | Duration (years) | | | | Development Cost ($M) | | | | Success Probability (%) | | | | Peak Sales ($M/yr) |
|---|---|---|---|---|---|---|---|---|---|---|---|---|---|
|  | Ph1 | Ph2 | Ph3 | Reg | Ph1 | Ph2 | Ph3 | Reg | Ph1 | Ph2 | Ph3 | Reg |  |
| Project 1 | 2 | 3 | 4 | 1 | 100 | 200 | 500 | 40 | 50 | 40 | 70 | 90 | 600 |
| Project 2 | -* | 2 | 3 | 1 | - | 150 | 300 | 40 | - | 30 | 60 | 90 | 400 |
| Project 3 | - | 2 | 4 | 1 | - | 150 | 400 | 40 | - | 30 | 50 | 90 | 700 |
| Project 4 | - | - | 3 | 1 | - | - | 500 | 40 | - | - | 70 | 95 | 400 |
| Project 5 | 2 | 2 | 3 | 1 | 70 | 150 | 300 | 40 | 60 | 30 | 60 | 90 | 200 |
| Project 6 | - | 3 | 4 | 1 | - | 200 | 500 | 40 | - | 30 | 60 | 90 | 1 000 |
| Project 7 | - | 1 | 3 | 1 | - | 100 | 300 | 40 | - | 30 | 50 | 90 | 400 |
| Project 8 | - | - | 3 | 1 | - | - | 400 | 40 | - | - | 60 | 90 | 300 |
| Project 9 | 2 | 3 | 3 | 1 | 100 | 150 | 500 | 40 | 60 | 30 | 70 | 90 | 1 300 |
| Project 10 | - | 2 | 4 | 1 | - | 100 | 300 | 40 | - | 40 | 70 | 90 | 800 |

*\* Note: Data for already completed phases are not relevant to the analysis, and not included in the table. For example, a project currently in Phase 2 does not have values in the table for Phase 1 duration, development cost and success probability.*

A PIT-plot has been produced for the Productivity Index metric, as described in previous sections, and the resulting graph is shown in Figure 3. The results of the illustrating example may suggest for the portfolio manager to consider the following findings:

- If the company has alternative investment opportunities or wants to improve return from a scarce development budget, Projects 1 and 5 might be candidates for termination. These projects occur at the bottom of the PIT-plot and the efficiency of the portfolio, as measured by the Productivity Index, would increase if these projects were excluded (i.e. the blue Exclusion bar has a positive value). These projects also have a limited positive upside, as indicated by relatively short green Success bars. For Project 5, the exclusion of the project would in fact have a positive impact on the portfolio that is even larger than the impact if the success of the project could be guaranteed.
- Projects 6 and 9 might be projects that the portfolio manager would want to pay additional attention to, in terms of mitigating risks of project failure and improving success probabilities. These projects have a potentially large positive impact on the portfolio metric (as indicated by large green Success bars).
- Projects 4 and 8 could be identified as projects to which the portfolio manager might give less attention, and the continuation according to current plans might be a prudent strategy. These projects are not candidates for exclusion, as they occur at the top of the PIT-plot, with negative values for the Exclusion bar. They do also have relatively small Success bars, hence the upside from project improvement efforts is lower than for some of the other projects.

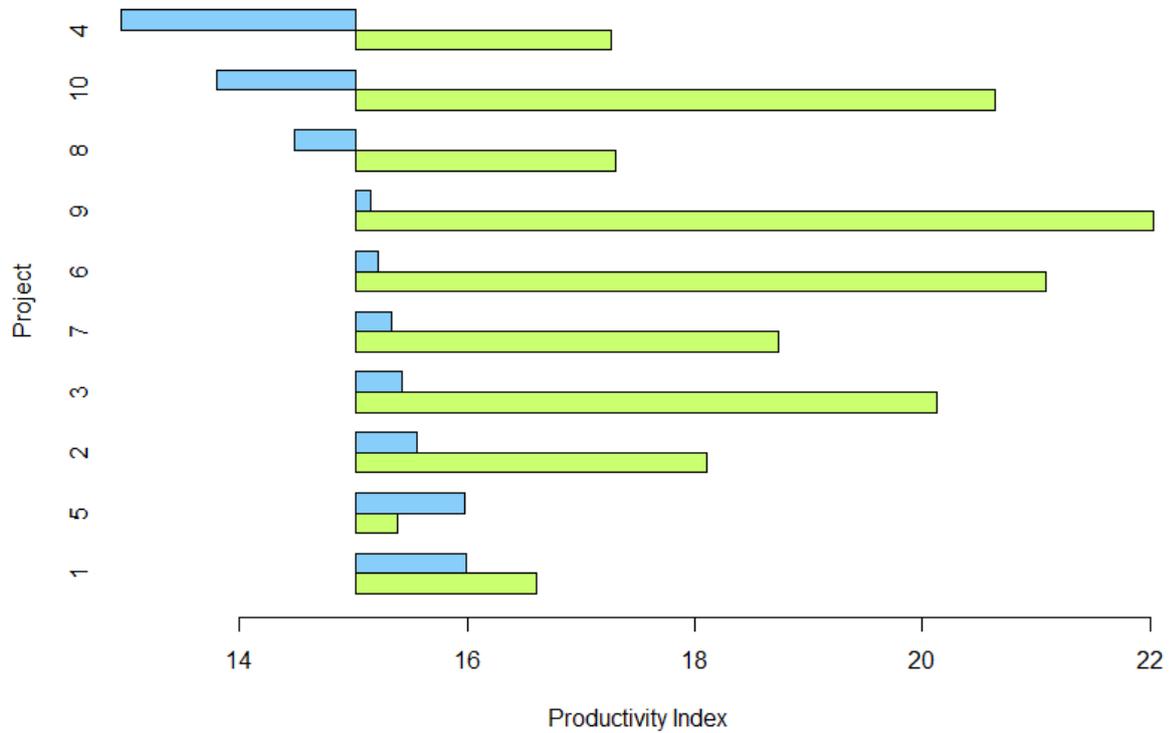

*Figure 3. The PIT-plot for the portfolio of ten projects included in the illustrating example, based on the Productivity Index metric.*

## 4. Discussion

In industries where project portfolio management (PPM) is vital, decision-making tools that enable a better understanding of portfolio dynamics are crucial (Jekunen, 2014). The Project Impact Tornado plot (PIT-plot) provides a novel approach to this challenge by shifting the focus from individual project sensitivities to a portfolio-level analysis, illustrating how each project impacts overall portfolio efficiency.

In cases where the projects in the portfolio are interdependent, decision-making and assessments of sensitivity should ideally take these dependencies into account. Such dependencies could arise in different ways. An example would be where two projects target the same market, and the success of one project might consequently reduce the value of another project. While we have in this article not explicitly addressed the concept of dependencies, the principles of the PIT-plot are valid also in these situations. The key requirement is that the models underlying the quantification of project and portfolio metrics are adequately capturing the essential dependencies.

The quantification of metrics for projects and portfolios requires forecasts and estimates to be made on a number of input parameters. These forecasts are inherently uncertain, implying that there are also substantial uncertainties in the resulting metrics. In the case of Monte Carlo simulation, an additional source of uncertainty would be the random variation inherent in the simulation procedure. These uncertainties propagate to the metrics visualized in the PIT-plot, obviously impacting the length of the bars but it may also impact the ordering of the projects in the plot. The size of the variability arising from the simulation procedure could theoretically be estimated, but the uncertainty of the input parameter estimates would remain unknown. While we have not in this paper explicitly addressed the issues of uncertainty, we recommend that the inherent uncertainties are acknowledged when applying and interpreting the results in a PIT-plot.

A main advantage of the PIT-plot lies in its ability to support portfolio selection and project prioritization. Traditional project ranking methods might favour individual metrics like Net Present Value (NPV) or Internal Rate of Return (IRR), but these approaches often miss the broader context. By capturing how each project affects the overall portfolio's efficiency, the PIT-plot offers a nuanced perspective. For example, if a portfolio manager is considering two projects with similar project metrics, the PIT-plot allows them to see which project contributes more effectively to the portfolio's Productivity Index, thereby helping them prioritize initiatives with better treatment efficacy.

The PIT-plot represents a valuable addition to the quantitative decision-making tools available to portfolio managers. Its unique structure allows managers to consider not only what happens if a project succeeds or fails but also how these outcomes impact the portfolio's overall performance. By focusing on productivity and efficiency, the PIT-plot complements traditional decision-making approaches by providing actionable insights that align project decisions with broader portfolio goals. When prioritizing in a portfolio of projects, relative measures (e.g. Return on Investment, Internal Rate of Return or Productivity Index) are particularly informative. Since these metrics relate the value of the project or portfolio to the investments needed, they explicitly measure the efficiency of the project investments. It is also for such relative measures that the PIT-plot is particularly informative. For additive metrics, managers can easily sort and rank projects based on simple calculations, making the PIT-plot less relevant in those cases. However, for metrics that require a portfolio-wide perspective on efficiency, the PIT-plot offers a clear advantage.

In summary, the PIT-plot is a flexible tool with wide applications across industries where PPM is critical. While its origins are in the pharmaceutical industry, its principles and benefits extend to any domain requiring portfolio-level decision-making. By offering a more holistic view of project impacts including situations with project dependencies and uncertainties, the PIT-plot can serve as a powerful decision-support tool for portfolio managers seeking to optimize their portfolios and achieve strategic objectives.

# References

Eschenbach TG. (1992) Spiderplots versus Tornado Diagrams for Sensitivity Analysis. *Interfaces* 22(6):40-46. https://doi.org/10.1287/inte.22.6.40

Farid M, Chaudhry A, Ytterstad M, Wiklund SJ. (2021). Pharmaceutical portfolio optimization under cost uncertainty via chance constrained-type method. *Journal of Mathematics in.Industry* **11**, 3. https://doi.org/10.1186/s13362-021-00099-3

Hu Q, Szmerekovsky J. (2017). Project Portfolio Selection: A Newsvendor Approach. *Decision Sciences*, 48: 176-199. https://doi.org/10.1111/deci.12214

Jekunen A. (2014). Decision-making in product portfolios of pharmaceutical research and development--managing streams of innovation in highly regulated markets. *Drug Design, Development and Therapy*. Oct 21;8:2009-16. doi: 10.2147/DDDT.S68579.

Lerch M and Spieth P, (2013). Innovation Project Portfolio Management: A Qualitative Analysis, *IEEE Transactions on Engineering Management*, 60:1, pp. 18-29, doi: 10.1109/TEM.2012.2201723.

Martinsuo M, Vuorinen L, Killen CP. (2024). Project portfolio formation as an organizational routine: Patterns of actions in implementing innovation strategy, *International Journal of Project Management*, 42: 4, https://doi.org/10.1016/j.ijproman.2024.102592.

Mohagheghi V, Mousavi SM and Mojtahedi M. (2020). Project Portfolio Selection Problems: Two Decades Review from 1999 to 2019'. *Journal of Intelligent & Fuzzy Systems*, 38:2, pp. 1675-1689, DOI: 10.3233/JIFS-182847

Raada NG, Shirazia MA and Ghodsypour SH. (2020) Selecting a portfolio of projects considering both optimization and balance of sub-portfolios. *Journal of Project Management*. 5;1–16.

Sampath S, Gel ES, Kempf KG, Fowler JW. (2022). A generalized decision support framework for large-scale project portfolio decisions. *Decision Sciences*. 53: 1024–1047. https://doi.org/10.1111/deci.12507

Schiffels S, Fliedner T, Kolisch R. (2018), Human Behavior in Project Portfolio Selection: Insights from an Experimental Study. *Decision Sciences*, 49: 1061-1087. https://doi.org/10.1111/deci.12310

Tiggemann RF, Dworaczyk DA, Sabel H. (1998) Project Portfolio Management: A Powerful Strategic Weapon in Pharmaceutical Drug Development. *Drug Information Journal*.32(3):813-824. doi:10.1177/009286159803200321

Vieira GB, Oliveira HS, de Almeida JA, Belderrain MCN. (2024). Project Portfolio Selection considering interdependencies: A review of terminology and approaches, *Project Leadership and Society*, Volume 5, 100115, ISSN 2666-7215, https://doi.org/10.1016/j.plas.2023.100115.

Wiklund SJ. (2019). A modelling framework for improved design and decision-making in drug development. PLoS ONE 14(8): e0220812. https://doi.org/10.1371/journal.pone.0220812

# Appendix 1: Tornado plot sensitivity analysis example

As a simple example to illustrate the use of Tornado plots for sensitivity analysis, we consider a manufacturing company, where total production cost is a function of fixed costs, variable production costs, and the quantity of items produced. For instance, assuming fixed costs of 300, variable costs of 10 per unit, and a production volume of 60 units, the resulting total cost would be 900. To conduct a sensitivity analysis, we adjust each cost component by ±10% to observe the effect on the total cost (see table below). This approach allows us to evaluate the degree to which changes in each parameter influence the overall costs.

| Scenario | Fixed Cost | Variable Cost | Items Produced | Total Cost |
|---|---|---|---|---|
| Base Case | 300 | 10 | 60 | 900 |
| Fixed Cost +10% | 330 | 10 | 60 | 930 |
| Fixed Cost -10% | 270 | 10 | 60 | 870 |
| Variable Cost +10% | 300 | 11 | 60 | 960 |
| Variable Cost -10% | 300 | 9 | 60 | 840 |
| Items Produced +10% | 300 | 10 | 65 | 950 |
| Items Produced -10% | 300 | 10 | 55 | 850 |

The outcome of this analysis can be visually represented using a Tornado plot, which effectively ranks the cost components according to their impact on total cost. In this particular example, as illustrated in the accompanying chart, variations in fixed costs exert the most significant influence on total costs, compared to changes in variable costs or production volumes.

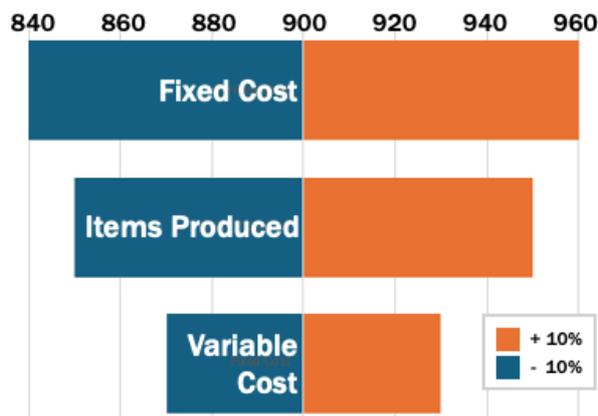